\documentclass[twocolumn,amsmath,amsfonts,amssymb,showpacs]{revtex4} %,nofootinbib
\usepackage{graphicx}
\def\beq{\begin{equation}}
\def\eeq{\end{equation}}
\def\bea{\begin{eqnarray}}
\def\eea{\end{eqnarray}}

\begin{document}

\title{Condensation of an ideal gas with intermediate statistics on the horizon}

\author{Somayeh Zare}
\affiliation{Department of Physics, Isfahan University of Technology, Isfahan, 84156-83111, Iran}
\author{Zahra Raissi}
\affiliation{Department of Physics, Isfahan University of Technology, Isfahan, 84156-83111, Iran}
\author{Hosein Mohammadzadeh}
\email{h.mohammadzadeh@gmail.com}
\affiliation{Department of Physics, University of Mohaghegh Ardabili , P. O. Box 179, Ardabil, Iran}
\author{Behrouz Mirza}
\email{b.mirza@cc.iut.ac.ir}
\affiliation{Department of Physics, Isfahan University of Technology, Isfahan, 84156-83111, Iran}
\affiliation{Research Institute for Astronomy and Astrophysics of Maragha (RIAAM), Maragha, Iran, P. O. Box 55134-441, Iran}

\begin{abstract}
We consider a boson gas on the stretched horizon of the Schwartzschild and Kerr black holes. It is shown that the gas is in a Bose-Einstein condensed state with the Hawking temperature $T_c=T_H$ if the particle number of the system be equal to the number of quantum bits of space-time  $ N \simeq {A}/{{\l_{p}}^{2}}$. Entropy of the gas is proportional to the area  of the horizon $(A)$ by construction. For a more realistic model of quantum degrees of freedom on the horizon, we should presumably consider interacting bosons (gravitons).  An ideal gas with intermediate statistics could be considered as an effective theory for interacting bosons. This analysis shows that we may obtain a correct entropy just by a suitable choice of  parameter in the intermediate  statistics.
\end{abstract}
\maketitle
%%%%%%%%%%%%%%%%%%%%%%%%%%%%%%%%%%%%%%%%%%%%%%%%%%%%%%%%%%%%%%%%%%%

%%%%%%%%%%%%%%%%%%%%%%%%%%%%%%%%%%%%%%%%%%%%%%%%%%%%%
\section{Introduction}
%%%%%%%%%%%%%%%%%%%%%%%%%%%%%%%%%%%%%%%%%%%%%%%%%%%%%
Many efforts have been made to find a statistical interpretation for the entropy of black holes. Starting from a theorem  proved by Hawking \cite{Hawking1,Hawking2}, Bekenstein conjectured that the entropy of black holes is proportional to the area of its event horizon ,
$S_{\textrm{BH}}={A}/{4{\l_{p}}^{2}},$
where, $A$ is the area of event horizon and $\l_{p}$ is the Planck length \cite{Bekenstein}.  Further  evidence for this was given in studies by  Hawking  \cite{Hawking3}.  Until 1995, no one was able to make a precise  calculation of black hole entropy based on a fundamental theory. The situation changed when
%A. Strominger and C. Vafa calculated
the right Bekenstein-Hawking entropy of a supersymmetric black hole was calculated by using a method
 based on D-branes \cite{vafa}. That calculation was followed by many similar computations of entropy of large classes of other extremal and near-extremal black holes. However, these methods cannot be applied to the more general case of nonsupersymmetric neutral black holes. Also, loop quantum gravity has yielded a detailed prescription for identifying microscopic quantum states
corresponding to an isolated horizon \cite{rovelli}.

Despite strong evidence for Bekenstein's conjecture, the physical nature of quantum mechanically distinct internal states has remained unknown. There is another different perspective for microstates of a Schwarzschild black hole based on earlier ideas of 't Hooft \cite{Hooft}, Susskind \cite{Susskind} and some other works \cite{zurek,sorkin}.  Gerlach tried to interpret the Hawking radiation as one produced by zero-point fluctuations on the surface of a star that collapsed to form  black holes. He concluded that the number $W_{zp}$ of zero-point fluctuation modes gives rise to the Hawking radiation of  freely evaporating Schwarzschild black holes is satisfied at $\ln W_{zp}\cong280S_{\textrm{BH}}$ \cite{Gerlach}. Also, according to York's proposal, Hawking radiation is produced by the black hole's quantum ergosphere of thermally excited gravitational quasinormal modes. He concluded that the number of ways $W_{qe}$  this quantum ergosphere can be excited and reexcited, during the evaporation of Schwarzschild hole into a surrounding radiation bath is satisfied at  $\ln W_{qe}\cong1.106 S_{\textrm{BH}}$ \cite{York}. It has been shown that one can\ consider the number of quantum mechanically distinct ways that the black holes could have been made by infalling quantum particles \cite{zurek}. The number of ways a Schwarzschild black hole of mass $M$ can be made by accretion of quanta from infinity ($r\gg2M$) will be a precise statistical explanation of $W$.

A great deal of effort has gone into resolving the puzzles of black hole thermodynamics and information loss. In this context, the idea of a stretched horizon arose as a useful tool for thinking about black holes. It originated as a classical description of black holes seen by an outside observers, but the concept was later borrowed to help give a consistent quantum mechanical interpretation of black hole physics \cite{thorne,susskind1}. It was also shown that the degrees of freedom of a stretched horizon can be viewed as a gas composed of quasi-particles \cite{lowe}. This simple picture makes manifest several universal properties of horizons, including the universal relationship between entropy and horizon area. Although a lot of research work  has been focused on explaining  the physics of horizon, our knowledge about the horizon degrees of freedom is still far from perfect. The simplest choice (not realistic) for the quantum degrees of freedom on the horizon is a noninteracting boson gas \cite{lowe}. We consider a simple model for the quantum degrees of freedom on the horizon. We assume that the horizon degrees of freedom are in a condensed state whose particle number is equal to the number of space-time quantum bits $N\simeq A/l_p^2$. For earlier studies on condensation on a  curved space-time see Ref. \cite{pathria2}.

   Since in a more realistic model we should consider a kind of interaction between bosons, in this paper, we explore an effective theory which is a version of brick-wall model with interacting bosons (gravitons). This model may give us some insight about the real quantum degrees of freedom on the horizon. The idea is similar to the brick-wall model introduced by 't Hooft \cite{Hooft}.
According to his derivation, the black hole entropy can be obtained by considering a quantum gas of scalar particles propagating outside the event horizon of the Schwarzschild black hole. The cutoff has been introduced for removing the divergence due to the infinite blueshift near the horizon. It is shown that the black hole entropy consists of mainly two parts: the leading order term as the standard Bekenstein-Hawking entropy and the logarithmic corrections to the black hole entropy.  In this context, another model, i.e. thin film model, has been introduced to improve the brick-wall model which can be applied to various black holes such as non-static black holes \cite{lix}.

 %It is not possible to make entropy of this gas $(T=T_H)$  equal to $A/4$ if we keep its particle number equal to the number of horizon degrees of freedom.
 Drawing upon the well-established fact that an interacting boson system could be represented by a deformed statistic \cite{scarfone}. We will consider a gas of particles with intermediate statistics as an (approximate) effective theory for the horizon degrees of freedom. For this purpose, two different intermediate statistics \cite{mohammad1,mohammad2} will be investigated in which a condensed state  of particles is possible. It is interesting that we may obtain entropy by determining the correct parameter in the statistics. The intermediate statistics is used here as a new tool to obtain the correct thermodynamic quantities. This toy model could give us some insights about the quantum structure of space-time. Our calculations indicate that the equipartition theorem on the stretched horizon should run as follows \cite{thooft, Suss3, pad2, Verlinde},
 \bea
 U_{H}\simeq N_{e}k_{B}T_{H}
 \eea
 where, $U_{H}$ and $T_{H}=T_c$ denote the internal energy and the Hawking temperature (local condensation temperature) on the stretched horizon, respectively, and $N_{e}=A/l_p^2$ is the   number of  quantum bits of space-time.\par
 The outline of this paper is as follows. In Sec. 2, the possibility of Bose-Einstein condensation on stretched horizon of Schwarzschild and Kerr-Newman space-time is investigated. In Sec. 3, and Sec. 4, two kinds of fractional statistics such as Polychronakos and q-deformed are introduced and some interesting properties of these intermediate statistics are explored. In Sec. 5, the thermodynamic quantities are obtained for an ideal fermion gas that is located near the horizon. Finally, in Sec, 6. the condensation on an arbitrary screen and the first law of thermodynamics are evaluated.

%%%%%%%%%%%%%%%%%%%%%%%%%%%%%%%%%%%%%%%%%%%%%%%%%%%%%
\section{Bose-Einstein condensation on the horizon}
%%%%%%%%%%%%%%%%%%%%%%%%%%%%%%%%%%%%%%%%%%%%%%%%%%%%%
First, we will evaluate the density of states in the background of a curved space-time. We consider a system with $N$ non-interacting particles described by a Hamiltonian $H(p_{i},q_{i})$. The system is confined to a specified volume and has an energy $E$. Therefore, $H(p_{i},q_{i})=E$ designates the phase space points located anywhere on the constant energy surface. We can start from the following quantities:
%%The volume of this surface is given by
%% \bea
 %%g_{N}(E)=\int dp_{i}dq^{i}\delta(E-H(p_{i},q_{i})),\nonumber
%% \eea
%%and the entropy of the system is $S(E)=k_{B}\ln g_{N}(E)$. As long as the particles are non-interacting, we can introduce $g(E)$ for a single particle.
\bea
\Gamma(E)=\int dp dq \ \Theta(E-H(p,q)),
\eea
where, $\Theta(x-x_{0})$ is the Heaviside step function and the density of the state is $g(E)=d\Gamma(E)/dE$. Now, for computing the volume of the phase space below an energy $E$ in the curved space-time, we have to use a covariant definition of energy. Such a definition of phase space volume in any static space-time is given by \cite{pad1}
 \bea
 \Gamma(E)=\int d^{d}x~d^{d}p~\Theta(E-\xi^{\alpha}p_{\alpha}),\label{DOS}
 \eea
where, $p^{a}$ is the four-momentum of the particle and $\xi^{a}\equiv(1,\textbf{0})$ is the killing vector of the static space-time. We construct the density of states   for a two-dimensional box in a static space-time. By using $p^{a}p_{a}=m^{2}$, one can write
 \bea
 \xi^{\alpha}p_{\alpha}=g_{00}p^{0}=g_{00}^{1/2}(m^{2}+\gamma^{\alpha\beta}p_{\alpha}p_{\beta})^{1/2}.\label{mom}
  \eea
where, $g_{00}$ and $\gamma_{\alpha\beta}$ are the metric elements of space-time and generally the line element is $ds^{2}=g_{00}dt^{2}-\gamma_{\alpha\beta}dx^{\alpha}dx^{\beta}.$ If we explore the system in various backgrounds the coefficient will be clear.

%%%%%%%%%%%%%%%%%%%%%%%%%%%%%%%%%%%%%%%%%%%%%%%%%%%%%
\subsection{Rindler space-time}
%%%%%%%%%%%%%%%%%%%%%%%%%%%%%%%%%%%%%%%%%%%%%%%%%%%%%
For a Rindler space-time, the line element is $ds^{2}=(1+\kappa x)^{2}dt^{2}-dx^{2}-dx_{\bot}^{2}$, where $\kappa$ is the surface gravity and Rindler observer perceives a horizon at $x=-1/\kappa$. Due to the quantum uncertainty about the position of any object, one cannot differentiate between an object that is Planck length distant from the horizon or one that has crossed the horizon. Therefore, we assume that the box reaches $x=-\kappa^{-1}+\l_{p}$. For a two dimensional box fixed  near the horizon, by using Eq. (\ref{mom}) one can evaluate
 \bea
 \Gamma(E)&=&\pi\int\sqrt{\gamma}d^{2}x_{\bot}(\frac{E^{2}}{\kappa^{2}{\l_{p}}^2}-m^{2})\nonumber\\
 &=&\pi A_{\bot}(\frac{E^{2}}{\kappa^{2}{\l_{p}}^2}-m^{2})
 \eea
Thus, the density of states is
 \bea
 g(E)=d\Gamma(E)/dE=2\pi A_{\bot}E/{\kappa^{2}{\l_{p}}^2},
 \eea
where, $A_{\bot}$ is the area of the box. We can work out the particle number using the derived density of states and the bosons mean occupation number,
 \bea
 N-N_{0}=\frac{2\pi A_{\bot}}{{\kappa^{2}{\l_{p}}^2}}\int\frac{E dE}{\frac{1}{z}e^{\beta E}-1}=\frac{2\pi A_{\bot}}{{\kappa^{2}{\l_{p}}^2}}(k_{B}T)^{2}Li_{2}(z),
 \eea
where, $Li_{n}(x)$ denotes the polylogarithm functions and $z=\exp(\beta\mu)$ is the fugacity of the boson gas. As we know, Bose-Einstein condensation occurs at $z=1$, and one can find the phase transition temperature at the constant particle density \cite{pathria},
 \bea
 {T_{c}}^{2}=\frac{{N \kappa^{2}{\l_{p}}^2}}{2\pi A_{\bot}{k_{B}}^{2}\zeta(2)},
 \eea
where, $\zeta(x)$ is the Reimann zeta function.  Hawking radiation temperature is $\beta_{H}=2\pi/\kappa$, and  one can, therefore, find a relationship such as that in (\ref{TCTH}) below to hold between phase transition temperature and the temperature of the horizon,
 \bea
 {T_{c}}^{2}={T_{H}}^{2}\frac{2\pi}{\zeta(2)}N\frac{{\l_{p}}^2}{A_{\bot}}.\label{TCTH}
 \eea

Now, a surprising result occurs when the particle number is equal to the number of quantum bits of space-time on the stretched horizon:
$N=(\frac{\zeta(2)}{2\pi}) ({A_{\bot}}/{{\l_{p}}^{2})}\simeq 0.26 ({A_{\bot}}/{{\l_{p}}^{2})}$. In this case,  the condensation temperature $T_{c}$ is equal to the Hawking temperature $T_{H}$,
 \bea
 {T_{c}}={T_{H}}\label{THTC}.
 \eea

  To gain more information about the quantum structure of space-time, the ideal boson gas may be viewed as  a probe. Therefore, we expect to find some information about thermodynamic quantities of an arbitrary two-dimensional screen ($r>2M$) by imposing $N\simeq A_{\bot}/{\l_{p}}^{2}$.  It should be interesting to investigate weak or strong interactions between particles of the gas or to use some other probes such as an ideal gas with fractional statistics. We will discuss these  elsewhere.
%%%%%%%%%%%%%%%%%%%%%%%%%%%%%%%%%%%%%%%%%%%%%%%%%%%%%
\subsection{Schwarzschild space-time}
%%%%%%%%%%%%%%%%%%%%%%%%%%%%%%%%%%%%%%%%%%%%%%%%%%%%%
In the following, we will consider the Schwarzschild space-time with the following metric,
 \bea
 ds^{2}&=&\left(1-\frac{2M}{r}\right)dt^{2}-\left(1-\frac{2M}{r}\right)^{-1}dr^{2}-r^{2}d\Omega^{2}.~~~~~~
 \eea
Obviously, the surface of constant energy is characterized by ${\gamma^{\alpha\beta}p_{\alpha}p_{\beta}}={E^{2}/g_{00}-m^{2}}$.
Now, consider a two-dimensional spherical box at a fixed radius and  containing non-interacting bosons in the  ultra-relativistic limit where the rest mass of particles is negligible. We derive some thermodynamic quantities  at stretched horizon ($r_{0}=2M+h$).
%% \begin{figure}[t]
 %%   % Requires \usepackage{graphicx}
%%   \center
 %%  \includegraphics[width=0.4\columnwidth]{HB2.eps}\\
 %%  \caption{(Color online) Horizon of a spherical 3 dimensional black hole and a two-dimensional spherical box, which located at a constant %% radius.}\label{BH}
 %%\end{figure}
 We will choose $h$ such that the proper length from the horizon is equal to the Planck length. Therefore, we will fix $h$ to be
 \bea
 \l_{p}=\int_{2M}^{2M+h}\frac{dr}{\sqrt{1-\frac{2M}{r}}}\approx2\sqrt{2hM}\label{planck}
 \eea
We evaluate $\Gamma(E)$ using Eq. (\ref{DOS}) as follows
 \bea
 \Gamma(E)&=&\int d^{2}x d^{2}p~\Theta\left(E-\sqrt{g_{_{00}}(m^{2}+p^{2})}\right)\nonumber\\
 &=&\pi\int\sqrt{\gamma}d^{2}x\left[\frac{E^{2}}{g_{00}}-m^{2}\right]\nonumber\\
 &=&\pi\Omega r_{0}^{2} \left[\frac{E^{2}}{1-\frac{2M}{r_{0}}}-m^{2}\right]
 \eea
and therefore, the density of states will be $g(E)=2\pi\Omega E{r_{0}^{2}}/{(1-\frac{2M}{r_{0}})}$,
where, $\Omega$ denotes the solid angle and $r_{0}$ is the radius where in the box is located. The particle number can be evaluated as follows
 \bea
 N-N_{0}=2\pi\Omega(k_{B}T)^{2}\frac{r_{0}^{2}}{1-\frac{2M}{r_{0}}}Li_{2}(z),
 \eea
 In this case, the phase transition temperature will be
 \bea
 {T_{c}}^{2}=\frac{N}{2\pi\Omega{k_{B}}^{2}\zeta(2)}\frac{1-\frac{2M}{r_{0}}}{r_{0}^{2}}.
 \eea
 If the box is located at near the horizon ($h\ll2M)$, we approximate ${[1-(2M/r_0)]}\approx{h}/{(2M)}$, and using Eq. (\ref{planck}), we will have
 \bea
 {T_{c}}^{2}=\frac{N}{2\pi\Omega{k_{B}}^{2}\zeta(2)}\frac{{\l_{p}}^{2}}{8M(2M)^{3}}.
 \eea
We notice that Hawking radiation temperature is given by the surface gravity, $\kappa={g^{\prime}}_{00}/2=1/2M$, and $A_{\bot}=\int\sqrt{\gamma}d^{2}x=\Omega r_0^{2}$.
Finally, one can work out a relation between BEC temperature on the stretched horizon and the Hawking radiation temperature of the black hole. As is expected, we obtain Eq. (\ref{THTC}) again.\par
For any spherically symmetric static space-time with the metric
 \bea
 ds^{2}=f(r)dt^{2}-\frac{1}{f(r)}dr^{2}-r^{2}d\Omega^{2},
 \eea
  the horizon ($r=r_{H}$) is defined by $f(r_{H})=0$ and $f^{\prime}(r_{H})=2\kappa$. It is straightforward to calculate the density of states $g(E)=2\pi{E A_{\bot}}/{f(r)}$ and the phase transition temperature will be obtained as in the preceding evaluation . We note that at locations near the horizon, the metric element can be approximated by $g_{_{00}}=f(r_0)\approx(r-r_{H})f^{\prime}(r_{H})=hf^{\prime}(r_{H})$.
If we use the proper length $\l_{p}$ from the horizon instead of $h$ by using
 \bea
 \l_{p}=\int_{{r_{_H}}}^{{r_{_H}}+h}\frac{dr}{\sqrt{f}}\approx2\sqrt{h/f^{\prime}(r_{H})}.
 \eea
 and the Hawking radiation temperature $T_H$ instead of $f^{\prime}(r_{H})$, by using $T_H=\frac{f^{\prime}(r_{H})}{4\pi k_B}$, we will have the following equation for $f(r_0)$, near the horizon:
 \bea
 f(r_{0})=4\pi ^{2}k_{B}^{2}l_{p}^{2}T_{H}^2 \label{ninety}
 \eea
Also, a similar relation will hold between  temperature of the phase transition and the Hawking temperature as in  Eq. (\ref{THTC}). One can evaluate the internal energy and entropy of the screen at the stretched horizon in the condensate state. Using the evaluated density of state and the well-known thermodynamic relations, we obtain
\bea
U_H&=&\int\frac{g(E)EdE}{\frac{1}{z}e^{\beta E}-1}=\frac{2}{3}\gamma_{0} (\frac{A_{\bot}}{{\l_{p}}^{2}})k_{B}T_{H}=\frac{2\gamma}{3}N_H k_{B}T_{c}\nonumber\\
A\ \ &=&k_{B}T\int g(E)\ln(1-ze^{-\beta E})dE = -\frac{\gamma}{3}N_H k_{B}T_{c}\nonumber\\
S_H&=&\frac{1}{T}(U-A)=\gamma N_H  k_{B}\simeq k_{B}\frac{A_{\bot}}{{\l_{p}}^{2}}
%\zeta -\frac{1}{3}\gamma_{0}k_{B}T_{c}\frac{A_{\bot}}{{\l_{p}}^{2}}=
\eea
where, $A$ is the Helmholtz free energy and $\gamma_{0}=\frac{3\zeta(3)}{2\pi}, \gamma=\frac{3\zeta(3)}{\zeta(2)}, $ and it is evident that $U=\frac{2}{3}ST_{H}$ \cite{Verlinde,pathria,hakim,pad2,hod2}. Therefore, the internal energy of the gas on the stretched horizon will be proportional to the mass of the black hole ($M\sim U\sim N_H k_{B}T_{H}$).
%%%%%%%%%%%%%%%%%%%%%%%%%%%%%%%%%%%%%%
\subsection{Kerr-Newman space-time}
%%%%%%%%%%%%%%%%%%%%%%%%%%%%%%%%%%%%%%
In this section, we calculate the thermodynamic quantities of a non-interacting boson gas at the stretched horizon of the Kerr-Newman space-time. The axisymmetric Kerr-Newman black hole solution is given by:
% consider the two-dimensional system in the shape of annular ring sub tending a solid angle $\Omega$ and
%with radius $r_{0}$, which is located near the horizon of the axisymmetric  Kerr-Newman space-time given by \cite{kang}:
\bea
ds^{2}=g_{tt}dt^{2}+2g_{t\phi }dtd\phi +g_{\phi \phi }d\phi  ^{2}+g_{rr}dr^{2}+g_{\theta \theta }d\theta^{2}.~~~
\eea
where
\bea
 &&g_{tt}=-\frac{\Delta -a^{2} \sin^{2}\theta }{\Sigma } ,\nonumber\\
 &&g_{t\phi }=-\frac{a \sin^{2}\theta \ (r^2+a^2-\Delta ) }{\Sigma },\nonumber\\
 &&g_{\phi \phi }=\frac{(r^2+a^2)^{2}- \Delta \ a^{2} \sin^{2}\theta}{\Sigma} \sin^{2}\theta ,\nonumber\\
 &&g_{rr}=\frac{\Sigma }{\Delta } ,\ g_{\theta \theta }=\Sigma
\eea
and
\bea
\Sigma =r^2+a^{2}\cos^{2} \theta  ,&& \Delta =r^{2}-2Mr+a^{2}+Q^2,~~
\eea
where, $M$, $a$, and $Q$ are mass, angular momentum per unit mass, and charge of black hole, respectively.
As  discussed in the preceding section, we need to calculate the phase space volume by the following relation:
\bea
\Gamma (E)=\int d\theta d\phi dp_{\theta }dp_{\phi }\label{Gamma (E)}
\eea
Here, we should note that the physical quantities must be measured locally in order for the entropy to be calculated  from the viewpoint of the observer near the horizon. \cite{israel}. Thus, the  observer measures local energy as  $\epsilon =E/\sqrt{-g^\prime _{tt}}$, where
\bea
-g^\prime _{tt}=-\frac{g_{tt}g_{\phi \phi }-g_{t \phi}^{2}}{g_{\phi \phi }}=\frac{\Delta sin^{2}\theta }{g_{\phi \phi }}\label{33}
\eea
And, due to the fact that the $p_{\phi }$ and $p_{\theta }$ integrals must be evaluated by calculating the area of the $p_{\phi }-p_{\theta }$ ellipse satisfying the following constraint condition:
\bea
\frac{p^2_{\theta }}{g_{\theta \theta }}+\frac{p^2_{\phi }}{g_{\phi \phi }}=\epsilon ^{2}-m^{2}
\eea
the phase space volume (\ref{Gamma (E)}) will be:
\bea
\Gamma (E)&=&\pi\int (\frac{E^{2}}{{-g^\prime _{tt}}}-m^{2})\sqrt{g_{\theta \theta }g_{\phi \phi }} d\theta d\phi \label{E}
\eea
To obtain the density of states, we have to differentiate with respect to the energy $(g(E)=\frac{d\Gamma (E)}{dE})$ and so we need the first term of the above relation. Using the Kerr-Newman metric and Eq. (\ref{E}), $\Gamma (E)$ can, therefore,  be rewritten as:
\bea
\Gamma (E)=2\pi^{2} E^2 \frac{1}{\Delta }\int\frac{[(r_0^2+a^2)^{2}-\Delta a^{2}sin^{2}\theta)]^{3/2}}{(r_0^2+a^2 cos^{2}\theta)}sin\theta d\theta ~~~
\eea
If we expand the numerator near the horizon, we will then have %which  $\Delta$ is small,
\bea
\Gamma (E)&=&2\pi^{2} E^2 \frac{1}{\Delta }\int\frac{(r_0^2+a^2)^3}{(r_0^2+a^2 cos^{2}\theta)}sin\theta d\theta\nonumber\\
&=&2\pi^{2} E^2 \frac{(r_0^2+a^2)^3}{\Delta }\left [\frac{2 \arctan(\frac{a}{r_0})}{ar_0}\right ]
 \eea
Finally, the density of states will be
\bea
g(E)=4\pi^{2} E\frac{(r_0^2+a^2)^3}{\Delta }\left [\frac{2 \arctan(\frac{a}{r_0})}{ar_0}\right ]\label{g(E)}
\eea
Using the derived density of states,  the mean number of particles will become:
\bea
N-N_0&=&\int\frac{g(E)dE}{z^{-1}e^{\beta E}-1}\nonumber\\
&=&4\pi^{2}\frac{(r_0^2+a^2)^3}{\Delta }(k_{B}T)^{2}Li_{2}(z)\left [\frac{2 \arctan(\frac{a}{r_0})}{ar_0}\right]~~~~\label{N-N_0}
\eea
Transition temperature takes place at z=1, when $N-N_0$ is practically identical with the total number of particles $N$; hence, the transition temperature becomes
\bea
T_c^2=\frac{N \Delta \ a \ r_0}{8\pi^{2} (r_0^2+a^2)^3 \arctan(\frac{a}{r_0})k_{B}^{2}\zeta (2)}
\eea
Near the horizon, $r_0=r_{+}+h$, will have:
 \bea
 T_c^2=\frac{N (h^2-2Mh+2r_{+}h)(r_{+}+h)^2}{8\pi^{2} (r_+^2+a^2)^3 \arctan(\frac{a}{r_+})k_{B}^{2}\zeta (2)} \label{T_c}
 \eea
where, $r_{+}=M+\sqrt{M^{2}-a^{2}-Q^{2}}$ is the location of the outer horizon of the Kerr-Newman black hole.  %$\Delta | _{r_+}=0$.
Also, the proper length to the horizon $l_{p}$ can be obtained from the following relation
\bea
l_{p}=\int_{r_{+}}^{r_{+}+R}\sqrt{g_{rr}}dr
\eea
Since the Kerr-Newman black hole is not spherically symmetric and the proper length $l_{p}$ must be constant, we should consider  $R=\frac{hr_+^2}{r_+^2+a^2 cos^{2}\theta}$, which is compatible with the results  reported elsewhere \cite{Ren}. In the following, we expend $\frac{1}{g_{rr}}=\frac{\Delta }{\Sigma }$  near the horizon
\bea
\frac{1}{g_{rr}}=\frac{2(r_{+}-M)}{(r_+^2+a^2 cos^{2}\theta)}(r-r_+)
\eea
So, for the minimal length from the horizon, we have
\bea
l_{p}=\int_{r_{+}}^{r_{+}+h}\sqrt{\frac{r_+^2+a^2 cos^{2}\theta}{2(r_{+}-M)}}\frac{dr}{\sqrt{r-r_+}}=\sqrt{\frac{2hr_+^2}{r_{+}-M}}~~~
\eea
Finally, $T_c$ that was obtained in Eq. (\ref{T_c}) can be expressed in terms of the Hawking temperature $T_H=\frac{\kappa }{2\pi k_{B}}=\frac{r_{+}-M}{2\pi k_{B} (r_{+}^2+a^2)}$ and the proper length $l_p$ from the horizon can be used instead of $h$:
\bea
T_c^2=T_H^2\frac{2{\pi}}{\zeta (2)}\frac{Nl_{p}^2}{A_\perp }\frac{a/r_{+}}{\arctan(\frac{a}{r_{+}})}
\eea
where, $A_\perp=\int\sqrt{g_{\theta \theta }g_{\phi \phi }} d\theta d\phi=4\pi (r^2+a^2)$  is the area of the two-dimensional spherical surface  located near the horizon.
 The transition temperature will be equal to the Hawking  temperature $(T_c=T_H)$ when the particle number is proportional to the following number of quantum bits of space-time on the stretched horizon:
\bea
N=\frac{\zeta(2)}{2\pi}\frac {A_{\bot}}{{\l_{p}}^{2}}(\frac{r_{+}}{a})\arctan(\frac{a}{r_{+}})\label{number}
\eea
For small values of  ${a}/r_{+}$, $\arctan({a}/r_{+})$ is approximately equal to  ${a}/r_{+}$  and we will have
\bea
N\simeq\frac{\zeta(2)}{2\pi}\frac {A_{\bot}}{{\l_{p}}^{2}}
\eea
which is exactly equivalent to Eq. (\ref{THTC}).  On the other hand, the largest value of $a/r_+$ (extremal limit) corresponds to the lowest value of N. It is interesting that by approaching to  a planck scale extremal black hole $(J= {M}^2)$ with $M=M_p=(\frac{\hbar c}{G^2})^{1/2}$ and  $J=\hbar$,  $N$ tends to the lowest possible positive integer $N\simeq 1$.

%%This limit is interesting and implies the largest possible value for the angular momentum of a black hole: $J_{max}=\frac{G^2 M^4}{\hbar c^2}$.

In the following, we will derive  the leading thermodynamic quantities. First of all, we evaluate the internal energy of the system by using the well-known thermodynamics relation.
 \bea
 U&=&\int \frac{g(E)EdE}{z^{-1}e^{\beta E}-1}\nonumber\\
 &=&8\pi^{2} \frac{(r_0^2+a^2)^3}{\Delta }\left [\frac{2 \arctan(\frac{a}{r_0})}{ar_0}\right ](k_{B}T)^{3}Li_{3}(z),~~~~\label{U1}
 \eea
Substituting (\ref{number}) in the above equation, and considering the fact that transition temperature occurs at $z=1$, we obtain the internal energy as follows
\bea
U=\frac{2}{3}\gamma Nk_{B}T_c\label{U22}
\eea
The Helmholtz free energy $A$ is given by the following formula
 \bea
 A&=&k_{B}T\int g(E)\ln(1-ze^{-\beta E})dE\nonumber\\
 &=&-4\pi^{2} \frac{(r_0^2+a^2)^3}{\Delta }\left [\frac{2 \arctan(\frac{a}{r_0})}{ar_0}\right ](k_{B}T)^{3}Li_{3}(z)\nonumber\\
 &=&-\frac{\gamma }{3}Nk_{B}T_c\label{Az=1},\label{A}
 \eea
 %%%By using Eq. (\ref{number}) and inserting $z=1$, we will have
 %%%\bea
%%% A=-\frac{\gamma }{3}Nk_{B}T_c\label{Az=1}
 %%%\eea
Using  (\ref{U22}) and (\ref{Az=1}), we can obtain the  entropy-area relationship of the stretched horizon,
 \bea
 S=\frac{1}{T}(U-A)=\gamma Nk_{B}
 =\frac{\gamma _{0}}{4}k_{B}\frac{A_{\bot}}{\l_{p}^2}(\frac{r_{+}}{a})\arctan(\frac{a}{r_{+}})~~~
 \eea
In the limit $a\rightarrow 0$, the above expression tends to  the corresponding expression  obtained in section $2$. Up to now, we have considered a noninteracting boson gas as a model for the quantum degrees of freedom on the horizon. This toy model is too simple and  we may improve it by assuming a gas of interacting bosons. In the next two sections, we will consider  a gas of particles with two kinds of intermediate statistics. These models could be considered as  effective theories of interacting particles \cite{scarfone}. They are more realistic and are also capable of showing certain aspects of the degrees of freedom on the horizon.
%%%%%%%%%%%%%%%%%%%%%%%%%%%%%%%%%%%%%%
\section{Polychronakos fractional statistics}
%%%%%%%%%%%%%%%%%%%%%%%%%%%%%%%%%%%%%%
 As a first approximation we could consider a free gas of gravitons to represent quantum degrees of freedom on the horizon. In a more realistic theory we should consider interaction between gravitons. A dense state of gravitons (bosons) is similar to an interacting boson gas in a curved background. For such interacting gas of bosons (gravitons) we assume that intermediate statistics is a useful effective theory. This means that interaction in a dense gas of gravitons (degrees of freedom on the horizon) might be formulated by a simple gas with intermediate statistics on a curved background. We wil consider this theory as a good approximation for a gas of gravitons. For a particular value of the parameter of the intermediate statistics we have a specific interaction between particles and by varying the parameter we actually change the strength of the interaction. The fractional statistics has been discussed as a generalization of Bose-Einstein and Fermi-Dirac statistics. There are some fractional exclusion statistics such as Haldane, Gentile, and Polychronakos \cite{Haldane,Gentile,Polychronakos}. In this section,  we will investigate  the fractional exclusion statistics that was  introduced by Polychronakos. The distribution function of Polychronakos fractional statistics  is of the following form:
\bea
n=\frac{1}{e^{\beta (\epsilon -\mu )}+(2k-1) }
\eea
where, $k=0$, $k=1$ and $0<k<1$ correspond to bosons, fermions, and Polychronakos fractional statistics, respectively. The thermodynamic geometry of the Polychronakos fractional statistics was studied in  \cite{mohammad1}. It was  shown that the thermodynamic geometry of Polychronakos fractional statistics could be divided into two different regions. For $k<0.5$, the thermodynamic curvature is positive and  condensation may occur  in this region as explored in \cite{mohammad1}. In the following, we investigate the Bose-Einstein condensation of   Polychronakos fractional statistics gas on the stretched horizon.
By using density of states for the general form of the metric given in section 2.2,
 $g(\epsilon)=\frac{dP(\epsilon)}{d\epsilon}=2\pi \Omega \epsilon \frac{r_{0}^{2}}{f(r_{0})}$, the particle number is given by:
\bea
N-N_0&=&\int n(\epsilon )g(\epsilon )d\epsilon\\
\nonumber &=& \frac{2\pi \Omega r_{0}^{2}}{f(r_{0})}\int\frac{\epsilon d\epsilon}{z^{-1}e^{\beta\epsilon}+(2k-1)}\\
\nonumber &=&\frac{2\pi A_\bot (k_{B}T)^{2}}{f(r_{0})}\ \frac{Li_{2}(z-2kz)}{(1-2k)}
\eea
Since occupation numbers cannot be negative or infinite, it is obvious that $0 \leq z(1-2k)\leq 1 $. Therefore, the phenomenon of  condensation occurs at $z(1-2k)=1$ and the phase transition temperature of the system is:
\bea
T_{c}^{2}&=&\frac{N f(r_{0})}{2\pi A_\bot k_{B}^{2}\zeta (2)}(1-2k).\label{28}
\eea
where, $\zeta (2)$ is the Riemann zeta function. Near the horizon, by substituting $f(r_{0})=4\pi ^{2}k_{B}^{2}l_{p}^{2}T_{H}^2$ in the above relation, we will have:
\bea
T_{c}^2=T_{H}^{2}\frac{2\pi }{\zeta (2)}\frac{Nl_{p}^{2} }{A_\bot }(1-2k).
\eea
Assuming that the particle number $N$ near the horizon is given by,
\bea
N=\frac{\zeta (2)}{2\pi}\frac{A_\bot }{l_{p}^{2}(1-2k)}.\label{30}
\eea
the condensation temperature $T_{c}$ will be equal to the Hawking temperature $T_{H}$, as in Eq. (\ref{THTC}).
To calculate the entropy of this system, we need the internal energy and Helmholtz free energy as given by:
\bea
\nonumber U&=&\int\frac{\epsilon g(\epsilon )d\epsilon}{z^{-1}e^{\beta\epsilon}+(2k-1)}\\
 &=&\frac{2\pi A_\bot (k_{B}T_c)^{3}}{f(r_{0})}\frac{\Gamma (3)Li_{3}(z-2kz)}{1-2k}\label{35}
 %&=&2Nk_{B}T_c\frac{\zeta (3)}{\zeta (2)}\\
\eea
\bea
\nonumber A&=&\frac{-k_{B}T}{2k-1}\int\ln (1+(2k-1)ze^{-\beta \epsilon })g(\epsilon )d\epsilon +\mu N\\
&=&-\frac{2\pi A_\bot (k_{B}T)^{3}}{f(r_{0})(1-2k)}[ Li_{3}(z-2kz)-\ln z Li_{2}(z-2kz)]~~~~~~~~
\eea
where, we have used the definition of fugacity $z\equiv e^{\beta \mu }$. If we insert $z(1-2k)=1$, for condensation, then we will obtain the relation below
\bea
U=2Nk_{B}T_c\frac{\zeta (3)}{\zeta (2)}&&
%\nonumber A&=&-\frac{8\pi A_\bot (k_{B}T_c)^{2}}{f(r_{0})(1-2k)}[k_{B}T_c \zeta (3)-\ln z \zeta (2)]\\
A=-N[k_{B}T_c \frac{\zeta (3)}{\zeta (2)}-\ln z]\label{33}
\eea
In the following, we can work out the entropy of the system by using Eqs. (\ref{30}), (\ref{33}):
\bea
 S=\frac{1}{T}(U-A)=N[3k_{B}\frac{\zeta (3)}{\zeta (2)}+\ln (1-2k)]=\gamma_{1} k_{B}\frac{A_\bot }{l_{p}^{2}}~~~~
\eea
where, $\gamma_{1}=\frac{1}{1-2k}[\frac{3\zeta (3)}{2\pi }+\frac{\zeta (2)}{2\pi}\ln (1-2k)]$. We can see for $k=0.4$, the entropy near the horizon will be $S=\frac{1}{4}\frac{A_\bot }{l_{p}^{2}}$. So, an ideal gas with  this kind of intermediate statistics may be considered as an effective theory (toy model) for quantum degrees of freedom on the horizon.

%%%%%%%%%%%%%%%%%%%%%%%%%%%%%%%%%%%%%%
\section{q-oscillators algebra}
%%%%%%%%%%%%%%%%%%%%%%%%%%%%%%%%%%%%%%
In this section, we will explore a system with $N$ relativistic and non-interacting q-deformed bosons as horizon degrees of freedom. In recent years, much attention has been directed to deformed statistics, such as fractional, Quon statistics, Anyon statisticas etc.
The theory of q-deformed bosons, or q-oscillators, was introduced based on  quantum groups with particular deformation parameters. A great deal of effort has been directed toward obtaining the thermodynamic and statistical properties of a q-deformed boson gas by Jackson derivative. These efforts have resulted in deriving both  the expressions of the physical quantities such as Bose-Einstein condensation temperature and the heat capacity of the system \cite{mohammad2}.
In this section, we will first introduce the q-deformed algebra defined by the following relations \cite{Lee,Ng,chaichian}:
\bea
[c,c]_k=[c^\dagger ,c^\dagger ]_k=0,  && cc^\dagger -kq^{k}c^\dagger c=q^{-N}\\
\nonumber [N,c^\dagger ]=c^\dagger ,   && [N,c]=-c
\eea
where, $[x,y]_k=xy-kyx$, with $k=1$ for q-bosons and $k=-1$ for q-fermions, and
$c^\dagger,c$  are the creation, annihilation operators and $N$ is the q-number operator, respectively, for which the following relations hold true:
\bea
c^\dagger c=[N],  && cc^\dagger =[1+kN]
\eea
where, $[..]$ stands for the q-basic number that is defined as:
\bea
[x]=\frac{q^{x}-q^{-x}}{q-q^{-1}}
\eea
Furthermore, in the  q-deformed structures, we need to the Jackson derivative (JD):
\bea
 D^{(q)}_{x}f(x)=\frac{f(qx)-f(q^{-1}x)}{x(q-q^{-1})}
\eea
In the  $q\rightarrow 1$ limit, JD reduces to the ordinary derivative.

%We consider a system with $N$ relativistic and non-interacting q-deformed bosons located in
%a statistic space-time with horizon. We merely know that the system is confined in a box with one
%dimension less than space.\par
The Hamiltonian of a system with $N$ relativistic and non-interacting q-deformed bosons has the following form:
\bea
H=\sum _{i}(\epsilon   _{i}-\mu )N_i
\eea
where, $\epsilon _{i}$ is the kinetic energy in the state i with the number operator $N_i$ and $\mu $ is the chemical potential.
The logarithm of the grand partition function for this system is given by:
\bea
\ln Z=-\sum _{i}\ln( 1-ze^{-\beta \epsilon _{i}})
\eea
where, $z=e^{\beta \mu }$ is the fugacity.
It is easy to show that the thermodynamic quantities in a q-deformed theory can be obtained  from JD derivative, rather than from the ordinary one.
So, the number of particles and internal energy can be obtained from the relations:
\bea
N=z D^{(q)}_{z}\ln Z\equiv\sum _{i}n_i, && U=-\epsilon_i D^{(q)}_{y_i}\ln Z\equiv\sum _{i}\epsilon_{i} n_{i}~~~~~~ \label{57}
\eea
where, $n_i$ is the mean occupation number that is obtained as follows \cite{Rubin}:
\bea
\nonumber[n_i]=\frac{-1}{\beta }\frac{\partial }{\partial \epsilon _i}\ln Z&=&\frac{-1}{\beta }\frac{\partial y_i}{\partial \epsilon _i}D_{y_{i}}\ln (1-kzy_{i})\\
 &=&\frac{1}{q-q^{-1}}\ln(\frac{z^{-1}e^{\beta \epsilon _i}-q^{-1}}{z^{-1}e^{\beta \epsilon _i}-q})\label{n_i}
\eea
We should note that the JD applies with respect to the $z=e^{\beta \mu }$ or $y_{i}=e^{-\beta \epsilon _i}$ that are in the exponential form.\par
With regard to the fact that  the spectrum of the particle states of the system in the thermodynamic limit, is almost a continuous one, the summation in Eq. (\ref{57}) can be replaced with the following integral.
\bea
\sum _{i}\longrightarrow \int g(\epsilon ) d\epsilon
\eea
where, $g(\epsilon )$ is the density of states. For calculating the thermodynamic quantities, we first need to obtain $g(\epsilon )$.
We shall be interested in the density of states of the general form of metric that was given in section 2.2,
 $g(\epsilon)=\frac{dP(\epsilon)}{d\epsilon}=2\pi \Omega \epsilon \frac{r_{0}^{2}}{f(r_{0})}.$

Now, we can work out the thermodynamic quantities using the mean occupation number in Eq.(\ref{n_i}), and the derived density of states:
\bea
\nonumber N-N_0&=&\int n(\epsilon )g(\epsilon )d\epsilon \\&=& \int\frac{1}{q-q^{-1}}\ln(\frac{z^{-1}e^{\beta \epsilon}-q^{-1}}{z^{-1}e^{\beta \epsilon }-q})  \frac{2\pi \Omega r_{0}^{2}}{f(r_{0})}\epsilon d\epsilon
\eea
By integrating by part, we obtain the following expression for $N-N_0$
\bea
N-N_0=\frac{2\pi \Omega r_{0}^{2}}{f(r_{0})}\frac{(k_{B}T)^2}{q-q^{-1}}H_{3}(z,q)\label{N-N_0}
\eea
where, $H_{n}(z,q)=Li_{n}(zq)-Li_{n}(zq^{-1})$ and $Li_{n}(x)$ denotes the polylogarithm function.\par
\noindent The internal energy for q-deformed statistic is defined by,
\bea
\nonumber U&=&\int \epsilon n(\epsilon )g(\epsilon) d\epsilon \\
\nonumber &=&\int\frac{1}{q-q^{-1}}\ln(\frac{z^{-1}e^{\beta \epsilon}-q^{-1}}{z^{-1}e^{\beta \epsilon }-q})  \frac{2\pi \Omega r_{0}^{2}}{f(r_{0})}\epsilon^2 d\epsilon\\
 &=&\frac{2\pi \Omega r_{0}^{2}}{f(r_{0})}\frac{(k_{B}T)^3}{q-q^{-1}}\Gamma (3) H_{4}(z,q)\label{UQ1}
\eea
We now proceed to calculate the Helmholtz free energy A that is given by
\bea
\nonumber A&=&k_{B}T\int g(\epsilon)\ln (1-z e^{-\beta \epsilon})d\epsilon +\mu N\\
 &=&k_{B}T\int \frac{2\pi \Omega r_{0}^{2}}{f(r_{0})} \epsilon \ln( 1-ze^{-\beta \epsilon})d\epsilon+\mu N
\eea
By using the prescription of the JD in the q-deformed algebra, and integration by part, we obtain the following expression  for A:
\begin{widetext}
\bea
\nonumber A&=&-(k_{B}T)^{3}\frac{2\pi \Omega r_{0}^{2}}{f(r_{0})}\frac{1}{q-q^{-1}}\int _{0}^{\infty }dx \frac{x^2}{2}\ln (\frac{z^{-1}e^{x}-q^{-1}}{z^{-1}e^{x}-q})+\mu N\nonumber\\
&=&-(k_{B}T)^{3}\frac{2\pi \Omega r_{0}^{2}}{f(r_{0})}\frac{1}{q-q^{-1}}[H_{4}(z,q)-(H_{3}(z,q)+N_{0})\ln z]\\
 &=&-(k_{B}T)^{3}\frac{2\pi \Omega r_{0}^{2}}{f(r_{0})}[h_{4}(z,q)-(h_{3}(z,q)+\frac{N_{0}}{q-q^{-1}})\ln z]\label{AQ}
\eea
\end{widetext}
where, $x=\beta \epsilon $ and $h_{n}(z,q)=\frac{1}{q-q^{-1}}H_{n}(z,q)$ while we know the deformed function $h_{n}(z,q)$, in the limit $q\rightarrow 1$, reduces to the Polylogarithm function $Li_{n}(z)$ for bosons and to the function $f_{n}(z)$ for fermions.
From the expressions for the  internal energy Eq.(\ref{UQ1}) and the Helmholtz free energy Eq.(\ref{AQ}), it is easy to show  that the entropy of the q-deformed gas has the following form:
\bea
S&=&\frac{1}{T}(U-A)\\
\nonumber &=&(k_{B}^{3}T^{2})\frac{2\pi \Omega r_{0}^{2}}{f(r_{0})}[3h_{4}(z,q)-(h_{3}(z,q)+\frac{N_{0}}{q-q^{-1}})\ln z]
\eea
 We now explore condensation of this q deformed Bose gas that is located in a static space-time with horizon. We can argue that there is an upper bound on the fugacity $z$ of the q-deformed boson gas where the deformed function $h_{n}(z,q)$ has its largest value.
 We pointed out that the upper bound on the fugacity can be obtained from two conditions: occupation number should be non-negative and its JD   must exist.
Therefore, the critical value of fugacity is given by \cite{Lavagno}:
\bea
   z_q=\left\{
         \begin{array}{cc}
           q^{2} & q<1 \\
           q^{-2} & q>1 \\
         \end{array}\right.\label{zq}
    \eea
We are now in a position to calculate the thermodynamic quantities such as the particles number (N), the internal energy (U), and finally the entropy (S) for the situation in which condensation occurs.
A similar analysis as in section 3 leads to the following particle number $N$.
\bea
N=\frac{2\pi \Omega r_{0}^{2}}{f(r_{0})}(k_{B}T_c)^2(h_{3}(z,q))_{max}
\eea
It is easy to show that the above relation reduces to:
\bea
T_{c}^2=\frac{Nf(r_{0})}{2\pi \Omega r_{0}^{2}k_{B}^2}\frac{1}{(h_{3}(z,q))_{max}}
\eea
By considering that near the horizon $f(r_{0})=4\pi ^{2}k_{B}^{2}l_{p}^{2}T_{H}^2$, the relation between $T_c$ and $T_H$ will be as follows:
\bea
T_{c}^2=\frac{2N\pi l_{p}^{2}T_{H}^2 }{ \Omega r_{0}^{2}}\frac{1}{(h_{3}(z,q))_{max}}
\eea
If one assumes that the particle numbers near the horizon is given by:
\bea
N=\frac{1}{2\pi }(h_{3}(z,q))_{max}\frac{A_\bot }{l_{p}^{2}}\label{70}
\eea
the condensation temperature $T_{c}$ will become equal to the Hawking temperature $T_{H}$, which is compatible with Eq. (\ref{THTC}).
Also we have the following expressions for internal energy and Helmholtz free energy:
\bea
U&=&2Nk_{B}T_c\frac{(h_{4}(z,q))_{max}}{(h_{3}(z,q))_{max}}\nonumber\\
A&=&-Nk_{B}T_c[\frac{(h_{4}(z,q))_{max}}{(h_{3}(z,q))_{max}}-\ln z_q]
\eea
We may derive the entropy of this system using Eq.(\ref{70}):
\bea
S&=&Nk_{B}[3\frac{(h_{4}(z,q))_{max}}{(h_{3}(z,q))_{max}}-\ln z_q]\\
\nonumber &=&\frac{k_{B}}{2\pi }\frac{A_\bot }{l_{p}^{2}}[3(h_{4}(z,q))_{max}-(h_{3}(z,q))_{max}\ln z]
\eea
Finally, we can see that, at $q=0.61$ (in the region $q<1$), and at $q=1.64$ (for the region $q>1$)
the entropy will be
$S=\frac{1}{4}\frac{A_\bot }{l_{p}^{2}}$, which is compatible with the Bekenstein-Hawking entropy. Thus, q-deformed statistics is successfully  considered as an  effective theory (toy model) for quantum degrees of freedom on the horizon.
%%%%%%%%%%%%%%%%%%%%%%%%%%%%%%%%%%%%%%
\section{Ideal fermions on the stretched horizon }
%%%%%%%%%%%%%%%%%%%%%%%%%%%%%%%%%%%%%%
In this section, we calculate the thermodynamic quantities of an ideal fermion gas on the stretched horizon.
The system has interesting properties in cases where the number of fermions is proportional to the number of space-time degrees of freedom, i.e. ${A}/{l_{p}^{2}}$

\noindent A similar calculation as in section 2.2 yields the following:
\bea
\nonumber N&=&\frac{2\pi A_{\perp}}{f(r_0)}\int_{0}^{\infty} \frac{E dE}{z^{-1}e^{\beta E}+1}\\&=&\frac{2\pi A_{\perp}}{f(r_0)}(k_{B}T)^{2}f_{2}(z)
=\frac{ A_{\perp} }{2\pi l_{p}^{2}}\frac{T^2}{T_{H}^2}f_{2}(z)\label{NF}
\eea
 where, the expansion of the metric element near the horizon, $f(r_{0})\simeq (r_{0}-r_{H})f'(r_{H})$, the proper length from the horizon, $l_{p}=\int _{r_{H}}^{r_{H}+h}\frac{dr}{\sqrt{f(r)}}\simeq 2\sqrt{\frac{h}{f'(r_{H})}}$ and Hawking radiation temperature $\beta_{H}=2\pi/\kappa$ have been used. \par
Unlike what happens in the Bose case, we do not encounter a phenomenon like Bose-Einstein condensation because of the Pauli exclusion principle. It is worth noting that  the system is assumed to be in thermal equilibrium with the horizon, i.e $T=T_H$, the particle numbers near the horizon will be almost equal to the number of space-time degrees of freedom on the horizon: $N=\frac{ A_{\perp} }{2\pi l_{p}^{2}}f_{2}(z)$.\par
\noindent The internal energy and Helmholtz free energy are defined as:
\bea
\nonumber U&=&\int\frac{\epsilon g(\epsilon )d\epsilon}{z^{-1}e^{\beta\epsilon}+1}
=\frac{2\pi A_\bot (k_{B}T)^{3}}{f(r_{0})}\Gamma (3)f_{3}(z)\\&=& k_{B} T_{H}f_{3}(z)\frac{ A_{\perp} }{\pi l_{p}^{2}}= 2Nk_{B}T_{H}\frac{f_{3}(z)}{f_{2}(z)}
\eea
\bea
\nonumber A&=&-k_{B}T\int g(\epsilon)\ln (1+z e^{-\beta \epsilon})d\epsilon +\mu N\\
\nonumber &=&-\frac{2\pi A_\bot (k_{B}T)^{3}}{f(r_{0})}\Gamma (3)f_{3}(z)+\mu N\\
&=&N k_{B}T(\ln z-\frac{f_{3}(z)}{f_{2}(z)})\nonumber\\
&=&k_{B}T_{H}\frac{ A_{\perp} }{2\pi l_{p}^{2}}(f_{2}(z)\ln z-f_{3}(z))
\eea
For the entropy of the system, we will have
\bea
  S=\frac{1}{T}(U-A)=N k_{B}(3\frac{f_{3}(z)}{f_{2}(z)}-\ln z)=\gamma_{2}k_{B}\frac{A_{\perp}}{l_{p}^{2}}
 \eea
where, $\gamma_{2}=\frac{1}{2\pi}(3f_{3}(z)-f_{2}(z)\ln z)$. Finally, we can see that at $z=0.44$, the derived entropy will be identical to the Hawking-Bekenstein entropy, i.e $S=\frac{k_{B}A_{\perp} }{4 l_{p}^{2}}$, in which the expansion of $f_{n}(z)$ was used for $z$ which is less than unity: $f_{n}(z)=z-\frac{z^2}{2^n}+\frac{z^3}{3^n}-...$). The internal energy of this system
will be $U\simeq 2Nk_{B} T_{H}$, meaning that the equipartition rule corresponds to a relativistic gas with two degrees of freedom as expected.
If the temperature of the gas is very low, the mean occupation number becomes
\bea
n=\frac{1}{z^{-1}e^{\beta E}+1}=\left\{
         \begin{array}{cc}
           1 & E< \mu _0 \\
           0 & E> \mu _0 \\
         \end{array} \right..
\eea
where, $\mu _0$ is the chemical potential of the system at $T=0$. So, at $T=0$, all single particle states are completely filled up to $E=\mu _0$, while all states with $E>\mu _0$ are empty. The limiting energy $\mu _0$ is referred to as the Fermi-energy of the system and is represented by  $\epsilon _f$. Near the horizon, one can obtain the particle number of this system, as follows
\bea
 N=\frac{2\pi A_{\perp}}{f(r_0)}\int_{0}^{\epsilon _f} E dE=
\frac{ A_{\perp} }{4\pi l_{p}^{2}}\frac{\epsilon _{f}^2}{k_{B}^2 T_{H}^2}
\eea
As already explained, near the horizon, the particle number of a fermion gas at Hawking temperature is equal to $N=\frac{ A_{\perp} }{2\pi l_{p}^{2}}f_{2}(z)$. Thus, we readily obtain the following interesting relation for Fermi-energy:
\bea
\epsilon _f=\sqrt{2 f_{2}(z)}k_{B}T_{H}
\eea
Although the geometry underlying general relativity is bosonic, our calculation indicates that fermions are also good candidate for fundamental structure of quantum bits of space -time. We would like to mention cooper pairs that are responsible for superconductivity. In lack of any experimental evidence we should also study such new scenarios.

%%%%%%%%%%%%%%%%%%%%%%%%%%%%%%%%%%%%%%%%%%%%%%%%%%%%%
\section{Condensation on an arbitrary screen ($r>2M$) and the first law of thermodynamics}
%%%%%%%%%%%%%%%%%%%%%%%%%%%%%%%%%%%%%%%%%%%%%%%%%%%%%
In discussing the properties of horizon, we have repeatedly run into the idea of the stretched horizon located a microscopic distance from the mathematical horizon. Similar to what we do for the horizon, we can assume an arbitrary screen ($r_{s}>2M$), so that the box of bose gas located at a Planck length away from it. For a recent study of thermodynamics on a maximally symmetric holographic screen which is relevant to our study see \cite{tian1}. It is worth to note that, due to the uncertainty in the position of any objects, one cannot really talk about the location of the box of gas $r_{0}$, that is located very close to the screen. Therefore, Like in the previous cases we consider $r_{0}=r_{s}+h$ to be very close to the screen such that $h/r_{s}\ll 1$. In the following section we again consider spherically symmetric static space-time with the metric
 \bea
 ds^{2}=-f(r)dt^{2}+\frac{1}{f(r)}dr^{2}+r^{2}d\Omega ^{2}.
 \eea
The distance of the particles from the screen can be explored by replacing $h$ instead of a Planck length $l_{p}$ which measures proper distance.
 \bea
 l_{p}&=&\int _{r_{s}}^{r_{s}+h}\frac{dr}{\sqrt{f(r)}}\nonumber\\
 &\simeq& \frac{2}{f'(r_{s})}\left(\sqrt{f(r_{s})+hf'(r_{s})}-\sqrt{f(r_{s})}\right),
 \eea
 Therefore, by using the derived equation, we can fix $h$ to be
  \bea
 h=\frac{l_{p}^{2}}{4}f'(r_{s})+l_{p}\sqrt{f(r_{s})},\label{h}
 \eea
We can also evaluate the relation between the phase transition temperature of the ideal two-dimensional boson gas located at an arbitrary distance from the horizon and a temperature equal to the Unruh temperature of the holographic screen \cite{Verlinde,wei,kono,tian,chen,mann,viss,wald}. It should be noted that we assume the Unruh-Verlinde temperature on a holographic screen as proposed in entropic force formulation of gravity \cite{Verlinde}. A boson gas can be used as a probe to explore equipartition theorem and entropy of  quantum bits of such screens. By using the derived density of state for the spherically symmetric static space-time, in section 2, we have,
 \bea
 T_{c}^{2}=\frac{1}{2\pi \zeta(2)}\frac{Nf(r_{0})}{k_{B}^{2}A_{\perp}},\label{Tcholographic}
 \eea
where $A_{\perp}=\int \sqrt{\gamma}d^{2}x=\Omega r_{0}^{2}$ denotes the area of the box. We note that, at near the screen the metric element can be approximated by $f(r_{0})\simeq f(r_{s})+(r_{0}-r_{s})f'(r_{s})=f(r_{s})+hf'(r_{s})$, also by using equations (\ref{h}),(\ref{Tcholographic}), we get
 \bea
 {T_{c}}= T_{u}+{\sqrt{f(r)}}T_{p}. \label{temper}
 \eea
where, $T_{u}=|f^{\prime}(r)|/4\pi$ denotes the Unruh temperature and $T_p ={1 }/{2\pi {k_{B}\l_{p}}}$ is the Planck temperature. It is evident that the above equation reduces to (\ref{THTC}) on the horizon. In other words, the condensation temperature on the holographic screen  is greater than the Unruh temperature and, therefore,  the quantum bits on the screen are in a condensed  state.

 We claim  that this  theory could be considered as a model for the quantum bits of space-time on a holographic screen.
Similar calculation for  an arbitrary holographic screen yields,

%%\cite{Hod}

\bea
U_u&=&\frac{2}{3}\gamma_{0} (\frac{A_{\bot}}{{\l_{p}}^{2}})k_{B}T_{u}({T_{u}\over {{T_{u}+{\sqrt{f(r)}}T_{p}}}})^2\simeq N_u k_{B}T_{u}({T_{u}\over T_c})^2\nonumber\\
S_u&=&\gamma N_u k_{B}({T_{u}\over {{T_{u}+{\sqrt{f(r)}}T_{p}}}})^2\simeq k_{B}\frac{A_{\bot}}{{\l_{p}}^{2}}({T_{u}\over {T_c}})^2. \label{entropy}
\eea

 In the above equation  $N_u k_{B}T_{u}=N_H k_{B}T_{H} \sim  Mc^2 $ is a constant. We propose that $(\ref{entropy})$ is a
 valid result for quantum bits of space-time on the holographic screen.\\
 Also we can calculate the thermodynamic quantities for an ideal gas with particles obeying Polychronakos fractional statistics located at an arbitrary holographic screen and we will see that the entropy for $k=0.4$ is as follows:
\bea
S&=&\frac{k_{B}A_\bot }{l_{p}^{2}}[\frac{3\zeta (3)}{2(1-2k)\pi }+\frac{\zeta (2)}{2(1-2k)\pi}\ln (1-2k)]({T_{u}\over T_c})^2\nonumber\\
&=& k_{B}\frac{A_\bot }{4l_{p}^{2}}({T_{u}\over T_c})^2
\eea

 The new equipartition theorem
 for an arbitrary holographic screen which is defined in $(\ref{entropy})$ leads to the following first law of thermodynamics.
 \bea
 dU=T dS- PdV + \mu dN
 \label{first}
 \eea

 \noindent where, $P\sim T_u (T_u/T_c)^2$ and $N \sim A_{\bot}/{\l_{p}}^{2}$. The presence of the last term is due to the fact that  the number of quantum bits on two neighboring  holographic screens are different, although for a single holographic screen $\mu=0$. It is simple to show that a part of $dU$ cancels out both $TdS$ and $PdV$ terms exactly. Another part of $dU$ is equal to $\mu dN $.  We propose a new
 physical energy interpretation for the quantum bits of space-time on  the holographic screen,  namely  $U_u= N_u k_{B}T_{u}({T_{u}\over T_c})^2=N_{e}k_{B}T_{u}=M ({T_{u}\over {T_{c}}})^2$. This physical energy on the holographic screen is too small to be detected at this time. This analysis leads to the following proposal:

{\it On an arbitrary  holographic screen the number of quantum bits of space-time in the ground $\{N_0 \}$ and excited $\{N_e=N_u-N_0 \}$ states are related by: $N_0=N_u (1- ({T_{u}\over {{T_{u}+{\sqrt{f(r)}}T_{p}}}})^2)$}.

\noindent This means that all quantum bits of an asymptotic Minkowski space-time $(T_u=0)$ are in the ground state. {\sl It should be noted that the definition of $U_u$ in  $(\ref{entropy})$ does not change the Newton's inverse square law if we use the entropic  interpretation of the gravitational force} \cite{Verlinde}. Another interesting result is that the horizon is a boundary where all quantum bits of space-time are going to be in the excited state. calculation of $ \mu$ from  $(\ref{first})$ yields:
\bea
 \mu\sim r ({T_{u}\over T_c})^2[\frac{d^2f}{dr^2}(1- {\sqrt f}\{\frac{T_p}{T_c}\})+\frac{8 \pi^2 k_B^2 }{{\sqrt f} }(\frac{T_p}{T_c})T_u^2]
 \label{mu}
 \eea
For $r >> 2 M$, $\mu\sim (1/r^7)$ is extremely  small and tends to zero asymptotically. It is simple to interpret this behavior for the asymptotically  flat region where
most quantum bits of space-time are in the ground state and little energy is needed to add a quantum bit to the excited states on  the holographic screen.  On the other hand, on the horizon, $\mu$ goes toward large numbers, which is also expected. All excited states are full on the horizon and a large amount of energy is needed to add a quantum bit to the holographic screen. It seems that for a holographic screen inside a black hole quantum bits of space-time should start filling quantum states with an energy gap from the ground state. Based on this interpretation, the singularity is not a physical point as the lowest possible energy for a  quantum bit of space-time goes to infinity at that point. We will investigate this interpretation more completely elsewhere.
%%%%%%%%%%%%%%%%%%%%%%%%%%%%%%%%%%%%%%%%%%%%%%%%%%%%%
\subsection{Entropy of the screen by directly counting the degrees of freedom}
%%%%%%%%%%%%%%%%%%%%%%%%%%%%%%%%%%%%%%%%%%%%%%%%%%%%%
One can obtain the entropy of the holographic screen at stretched horizon  by counting the likely microstate. We can assume that the system has $G=A_{\bot}/l_p^2$ states to be occupied and $N$ quantum bits of space-time have to be distributed among these states. Also, we notice that at the condensate state $N=A_{\bot}/l_p^2=G$. Therefore, the number of distinct ways in which the $N$ identical and indistinguishable particles (quantum bits of space-time) can be distributed among the $G$ excited states in the thermodynamic limit will be
 \bea
 W=\frac{(G+N-1)!}{N!(G-1)!}\simeq\frac{(2N)!}{(N!)^{2}}\nonumber
 \eea
and, therefor, the entropy can be worked out  as a function of $N$ or equivalently $A_{\bot}/{{\l_{p}}^{2}}$  as follows
 \bea
 S&=&k_{B}\ln (W)\simeq a N+b\ln(N)+c\frac{1}{N}+d\frac{1}{N^3}+\cdots\nonumber\\
 &\simeq&a\frac{A_{\bot}}{{\l_{p}}^{2}}+b \ln({\frac{A_{\bot}}{{\l_{p}}^{2}}})+c{\frac{{\l_{p}}^{2}}{A_{\bot}}}+d {(\frac{{\l_{p}}^{2}}{A_{\bot}})^{3}}+\cdots,
 \eea
where, $a,b,\dots$ are constants. The above result is consistent with the known result that a massless scalar filed  yields logarithmic corrections to the entropy \cite{man}.
%For an arbitrary holographic screen, the system has $G=N_u-N_0=N_u(T_u/T_c)^2=N_e$  excited states and the effective number of particles (number of excited quantum bits of space-time)  is equal to $N_u(T_u/T_c)^2$ and so entropy is proportional to $(\ref{entropy})$.
It follows that in a Minkowski space where  all quantum bits of space-time are in the ground state (according to our proposal), the entropy is zero \cite{Bek2}. In this way, we may obtain  the  Bekenstein-Hawking entropy as well as its corrections which are  somewhat universal in theories of quantum gravity such as  loop quantum gravity and  string theory  \cite{ghosh,sen}.
%%%%%%%%%%%%%%%%%%%%%%%%%%%%%%%%%%%%%%%%%%%%%%%%%%%%%
\subsection{Boson gas as a fast scrambler}
%%%%%%%%%%%%%%%%%%%%%%%%%%%%%%%%%%%%%%%%%%%%%%%%%%%%%
It is simple to show that the time scale for a particle to diffuse over the entire physical system with temperature $T$ satisfies  $tT\sim N^{2/d}l_{0}\bar{p}$, where $N$ is the number of particles, $\bar{p}$ is the average momentum of a particle in thermal equilibrium and $l_{0}$ is length scale of the system, which for a dense gas can be the separation of the molecules or their size. In the case of strongly correlated quantum systems, the diffusion and scrambling times are of the same order. If the total entropy of the gas is in the order of the number of molecules, than the scrambling time is $t_{\ast}T=S^{2/d}\hbar$ , where $d$ is the dimension of the system. One might think that in the real world $t_{\ast}T$ never grows more slowly than $S^{2/3}$ \cite{sus3}. However, assuming that the Stefan-Boltzmann law gives the rate at which the stretched horizon emits energy in the outgoing Hawking radiation ($\frac{dU}{dt}\sim \frac{A_{\bot}}{{\l_{p}}^{2}}{T_{H}}^{2})$, where $t$ is the Schwarzschild time coordinate,  we will, therefore, have,
 \bea
 \frac{dS}{dt}\simeq T_{H}\frac{A_{\bot}}{{\l_{p}}^{2}}\sim T_{H}S
 \eea
Now, one can obtain a consistent scrambling time  $t_{\ast}\sim\beta\ln S$,  where $\beta$ is the inverse Hawking temperature.  It is remarkable that for a black hole, the scrambling time satisfies $t_{\ast}T\sim\hbar\ln S$.

%%%%%%%%%%%%%%%%%%%%%%%%%%%%%%%%%%%%%%%%%%%%%%%%%%%%%
\section{Concluding remarks}
%%%%%%%%%%%%%%%%%%%%%%%%%%%%%%%%%%%%%%%%%%%%%%%%%%%%%
The possibility of Bose-Einstein condensation for a boson gas on the stretched horizon of the Schwartzschild and Kerr-Newman space-times was investigated and it was found that in cases where the particle number of the system is proportional (equal) to the number of quantum bits of space-time $ N \simeq {A}/{{\l_{p}}^{2}}$, the gas will be in the condensed state with the Hawking temperature $T_c=T_H$. This condensed state represents a kind of brick-wall model and can be used to count horizon degrees of freedom. This finding is in certain ways similar to those of  a recent study about condensation and BTZ black holes \cite{vaz}. In a recent paper Dvali and Gomez (DG) \cite{Dvali} proposed that black holes are nothing but a Bose-Einstein condensate of gravitons. These are some similarities between our and DG proposal. For example the size of condensate $A\simeq N{\l_{p}}^{2}$ in our approach which is similar to $l\simeq \sqrt{N}\l_{p}$ in \cite{Dvali} and DG proposal for
the critical energy is $E_c=\frac{\sqrt{N}\hbar}{\l_{p}}$ ($l\simeq \sqrt{N}\l_{p}$ ) which is exactly similar to ours $U_c\simeq Nk_BT_H\simeq \frac{N\hbar}{l}=\frac{\sqrt{N}\hbar}{\l_{p}}$.
We may use a gas of interacting bosons as a theory to approximate horizon degrees of freedom. As intermediate statistics could represent interacting boson systems (gravitons), a gas of particles with intermediate statistics was considered as an effective theory of horizon degrees of freedom. The entropy of the interacting gas was regularized by choosing the correct parameter of the statistics. We may consider this interacting boson (graviton) gas as an effective coarse-grained theory for the quantum structure of space-time. Another finding of the present study implies a new form of equipartition theorem for holographic screens, which reduces to the known  formula on the horizon. Our proposal indicates a finite entropy for the holographic screen that does not diverge at infinity. It will be interesting to see if the above mentioned results could be obtained by other methods.

\textbf{Acknowledgments}:
We would like to thank R. Casadio for useful comments. This work has been supported financially by the Research Institute for Astronomy and Astrophysics of Maragha (RIAAM) under research project No. 1/2359.

%%%%%%%%%%%%%%%%%%%%%%%%%%%%%%%%%%%%%%%%%%%%%%%%%%%%%

%%%%%%%%%%%%%%%%%%%%%%%%%%%%%%%%%%%%%%%%%%%%%%%%%%%%%%%%%%%%%%%%%%%%%%%%%%%%

\end{document}